\documentclass[prl,aps,twocolumn,amsmath,amssymb,amsfonts,floatfix,superscriptaddress,longbibliography]{revtex4-1}

\usepackage{graphicx}
\usepackage{dcolumn}
\usepackage{bm}
\usepackage{mathrsfs,subcaption}
\usepackage{epstopdf}

\usepackage{mathtools} 

\usepackage[space]{grffile}

\usepackage{makecell} 

\graphicspath{{./figs/}}

\usepackage[breaklinks=true,colorlinks=true,linkcolor=blue,urlcolor=blue,citecolor=blue]{hyperref}

\usepackage{color}

\def\rd{\text{d}}

\def\bk{{\bf k}}
\def\br{{\bf r}}
\def\rdbr{\rd{\bf r}}
\def\bR{{\bf R}}

\def\cF{{\cal F}}
\def\cFex{{\cal F}_{\text{ex}}}
\def\cUext{U}

\begin{document}
\title{Density Distribution in Soft Matter Crystals and Quasicrystals}
\author{P. Subramanian}
\affiliation{Mathematical Institute, University of Oxford, Oxford OX2 6GG, United Kingdom}
\author{D.J. Ratliff}
\affiliation{Department of Mathematical Sciences and Interdisciplinary Centre for Mathematical Modelling, Loughborough University, Loughborough LE11 3TU, United Kingdom}
\affiliation{Department of Mathematics, Physics and Electrical Engineering, Northumbria University, Newcastle upon Tyne NE1 8ST, United Kingdom}
\author{A.M. Rucklidge}
\affiliation{School of Mathematics, University of Leeds, Leeds LS2 9JT, United Kingdom}
\author{A.J. Archer}
\affiliation{Department of Mathematical Sciences and Interdisciplinary Centre for Mathematical Modelling, Loughborough University, Loughborough LE11 3TU, United Kingdom}

\begin{abstract}
The density distribution in solids is often represented as a sum of Gaussian
peaks (or similar functions) centred on lattice sites or via
a Fourier sum. Here, we argue that representing instead the \emph{logarithm} of the density
distribution via a Fourier sum is better. We show that truncating such a representation after
only a few terms can be highly accurate for soft matter crystals. For quasicrystals, this
sum does not truncate so easily, nonetheless, representing the density profile in this
way is still of great use, enabling us to calculate the phase diagram
for a 3-dimensional quasicrystal forming system using an accurate non-local density functional theory.
\end{abstract}


\maketitle



The form of the average (probability) density distribution $\rho(\br)$ of particles in
crystalline and quasicrystalline solids depends crucially on various factors
such as temperature, pressure and the nature of the particle interactions. Many
important material properties are in turn sensitively related to the form of
$\rho(\br)$. For example, the Lindemann criterion~\cite{hansen,
singh1991density, lowen1994melting} identifies the melting of a crystal in
terms of the {\color{black} widths} of the peaks in $\rho(\br)$, which depend sensitively on
the distance in the phase diagram
between the current state and where solid--liquid phase coexistence occurs.


It has been known for some time that in a uniform solid, $\rho(\br)$~can be
represented well by sums of Gaussian peaks centred on the lattice 
sites~\cite{hansen, singh1991density, lowen1994melting, tarazona2000density},
i.e.,
 \begin{equation}
 \rho(\br)=\sum_l n\left(\frac{\alpha}{\pi}\right)^{3/2}e^{-\alpha(\br-\bR_l)^2},
 \label{eq:sum_of_gaussians}
 \end{equation}
where $\alpha$ controls the peak widths and
$\{\bR_1,\bR_2,\dots\}=\{\bR_l\}$ is the set of vectors of the
lattice sites in the solid. For a crystal these are the set of lattice
vectors {\color{black} and} $n$~is the average number of particles per lattice site.
If the particles have a hard core {\color{black}then} $n\leq1$, but for the
soft penetrable particles which model
polymeric molecules in solution
\cite{likos2001effective, Mladek2006PRL, lenz2012microscopically} considered here, $n>1$. The Gaussian
form~\eqref{eq:sum_of_gaussians} and its anisotropic generalisations ~\cite{singh1991density, lowen1994melting, tarazona2000density} are fairly accurate deep in a crystal phase, but are less accurate close to melting. 


{\color{black} The other standard representation, due to its periodicity,}
is to express~$\rho(\br)$ as a Fourier sum~{\color{black}\cite{hansen, singh1991density}}:
 \begin{equation}
 \rho(\br)=\sum_j{\hat\rho}_j\exp(i \bk_j\cdot\br),
 \label{eq:sum_of_fourier}
 \end{equation}
where $\{\bk_j\}$ is the set of reciprocal lattice vectors (RLVs) for
the crystal, including~$\bk=0$, and~${\hat\rho}_j$ are the Fourier coefficients.
For example, in a simple cubic crystal, these wavevectors form a cubic lattice,
and the smallest non-zero wavenumber is related to the size of the unit cell.


Both the representations above become more involved when considering
quasicrystals (QCs). These have the spatial order of crystals but they lack
periodicity, so in QCs the set of vectors $\{\bR_l\}$ is aperiodic and
Eq.~\eqref{eq:sum_of_gaussians} needs to be modified to allow the heights and
widths of the peaks to vary in space, replacing $(n,\alpha)$ by
$(\{n_l\},\{\alpha_l\})$. The representation in
Eq.~\eqref{eq:sum_of_fourier} can still be used for QCs, {\color{black}with the
RLVs} indexed by up to six integers~\cite{walter2009crystallography,
jiang2018icm}.
 

The density peaks of a solid can be rather sharp, so the
Fourier sum representation ~\eqref{eq:sum_of_fourier} requires a large number of
terms to be accurate. 
Here, we advocate the following alternative ansatz as being more
useful and accurate than either~\eqref{eq:sum_of_gaussians}
or~\eqref{eq:sum_of_fourier} for crystal and QC density distributions:
 \begin{equation}
 \rho(\br)=\rho_0\exp\bigg[\sum_j{\hat\phi}_j\exp(i \bk_j\cdot\br)\bigg],
 \label{eq:exp_sum_of_fourier}
 \end{equation}
namely, we represent the \emph{logarithm} of the density as a Fourier sum over
the RLVs, with Fourier coefficients~${\hat\phi}_j$ and $\rho_0$~an
arbitrary reference density.

The advantage of representation~\eqref{eq:exp_sum_of_fourier} is that it excels
both deep in the crystalline region of the phase diagram, where
\eqref{eq:sum_of_gaussians} is accurate, and also close to melting,
where the peaks broaden and \eqref{eq:sum_of_fourier}
becomes viable. We show below that for the soft matter systems considered
here, retaining only a few terms in the sum in~\eqref{eq:exp_sum_of_fourier}
can be remarkably accurate in both regimes. In~\cite{archer2019deriving} we
showed that simply retaining wavenumber zero and one other wavenumber
in~\eqref{eq:exp_sum_of_fourier} quantitatively agrees with a fully resolved
representation of the density in lamellar phases,
both near and far from melting.
{\color{black} We show here that minimal} extra effort is required for crystals such as face-centered
cubic (FCC) and body-centered cubic (BCC), although it transpires that more effort is
needed for dodecagonal QCs and icosahedral quasicrystals~(\hbox{IQCs}).
 

In what follows, we explain the procedure to determine $\rho(\br)$
in the framework of density functional theory~(DFT)~\cite{evans1979nature, Evans92, hansen}.
We show that a severely truncated and simplified ansatz based
on~(\ref{eq:exp_sum_of_fourier}) allows for an accurate and efficient
determination of the 3-dimensional (3D) density distributions, and compares well with
existing results~\cite{Mladek2006PRL,pini2015unconstrained}. Additionally, we
show how this strongly nonlinear theory~(SNLT) enables
efficient computation of the phase diagram in a system that is capable of forming
both 3D crystals and~IQCs.


The central quantity in DFT is the Helmholtz free energy, expressed as a functional of the
density:
 \begin{equation}
 \cF[\rho] = k_BT \int \rho\left(\ln(\Lambda^3\rho) - 1\right)\rdbr
             + \cFex[\rho] + \int\rho\cUext\rdbr.
 \end{equation}
The first term is the ideal-gas contribution, {\color{black}with $\Lambda$~the thermal
de Broglie wavelength, $k_B$ Boltzmann's constant and temperature $T$}. $\cFex$~is the excess contribution due to
particle interactions and the third term is from any
external potential~$\cUext(\br)$. We set $\cUext=0$, as we are
interested only in bulk behaviour.
The equilibrium density profiles minimize the grand potential
$\Omega[\rho]=\cF[\rho] - \mu\int\rho\rdbr$, where $\mu$~is the chemical
potential, and thus satisfy the Euler--Lagrange equation
 \begin{equation}
 \ln(\Lambda^3\rho) - c^{(1)}[\rho] - \beta\mu = 0,
 \label{eq:eulerlagrange}
 \end{equation}
{\color{black} where} {\color{black} $\beta=(k_BT)^{-1}$} and $c^{(1)}[\rho]\equiv-\beta\delta\cFex/\delta\rho$ {\color{black}is} the one-body direct
correlation function~\cite{evans1979nature, Evans92, hansen}. {\color{black}Taking a functional derivative of \eqref{eq:eulerlagrange} w.r.t.\ $\rho$ and then integrating again yields a formally exact expression for $c^{(1)}[\rho]$ that can be rearranged to give}
{\color{black} 
\begin{equation}
\rho(\br)=\rho_0\exp\bigg[\int\rdbr'(\rho(\br')-\rho_0)\hspace{-1mm}\int_0^1 \hspace{-1.5mm}\textrm{d}\lambda c^{(2)}(\br,\br';\rho_\lambda)\bigg],
\label{eq:exact}
\end{equation}
which is obtained by thermodynamic integration (details in the supplementary information \cite{suppmat})
along a sequence of states with profiles
$\rho_\lambda\!\!\!=\!\!(1-\lambda)\rho_0\!+\!\lambda\rho$. Here, $c^{(2)}(\br,\br';\rho_\lambda)$
is the pair direct correlation function for the inhomogeneous systems along this path
\cite{hansen, singh1991density}. For a system with interparticle pair-potential $v(r)$,
the function $c^{(2)}(\br,\br';\rho_\lambda)\sim-\beta v(|\br-\br'|)$ for large
$|\br-\br'|$ and is finite for all $(\br,\br')$. Thus,
the spatial integral inside the exponential in Eq.~\eqref{eq:exact} 
has the effect of smearing the sharp peaks in $(\rho(\br)-\rho_0)$ and so is
more slowly varying than the density, meaning it can be represented accurately
via a Fourier sum with fewer terms. The exponential of this smooth function is then the sharply peaked density.}


The first term in~\eqref{eq:eulerlagrange} {\color{black}provides further} motivation for the ansatz~\eqref{eq:exp_sum_of_fourier}. 
Substituting $\rho=\rho_0\exp(\phi)$ into~\eqref{eq:eulerlagrange}
(without assuming the system is periodic), we obtain
 \begin{equation}
 \ln(\Lambda^3\rho_0) + \phi - c^{(1)}[\rho_0e^{\phi}] - \beta\mu = 0.
 \label{eq:eulerlagrangephi}
 \end{equation}
Fourier transforming gives
 \begin{equation}
 {\hat\phi} - \widehat{c^{(1)}[\rho_0e^{\phi}]} - \beta\mu^*\delta(\bk) = 0,
 \label{eq:eulerlagrangeft}
 \end{equation}
where the circumflex denotes the Fourier transform. We 
define $\beta\mu^* = \beta\mu-\ln(\Lambda^3\rho_0)$,
i.e.\ the chemical potential with a constant subtracted,
and $\delta(\bk)$ is a Dirac delta function.
When the system is periodic, we can replace the Fourier transforms
in~\eqref{eq:eulerlagrangeft} by Fourier sums and the Dirac delta becomes the
Kronecker delta $\delta_{\bk,0}$. With the
ansatz~\eqref{eq:exp_sum_of_fourier}, the unknown Fourier
amplitudes~$\hat{\phi}_j$, are found by solving
Eq.~\eqref{eq:eulerlagrangeft}. The advantage of this is that we are working
\emph{with}, rather than against, the physics, and fewer modes
in~\eqref{eq:exp_sum_of_fourier} are needed to resolve $\rho(\br)$ accurately.
\begin{figure*}[t]
\centering{
\includegraphics[width=0.95\columnwidth]{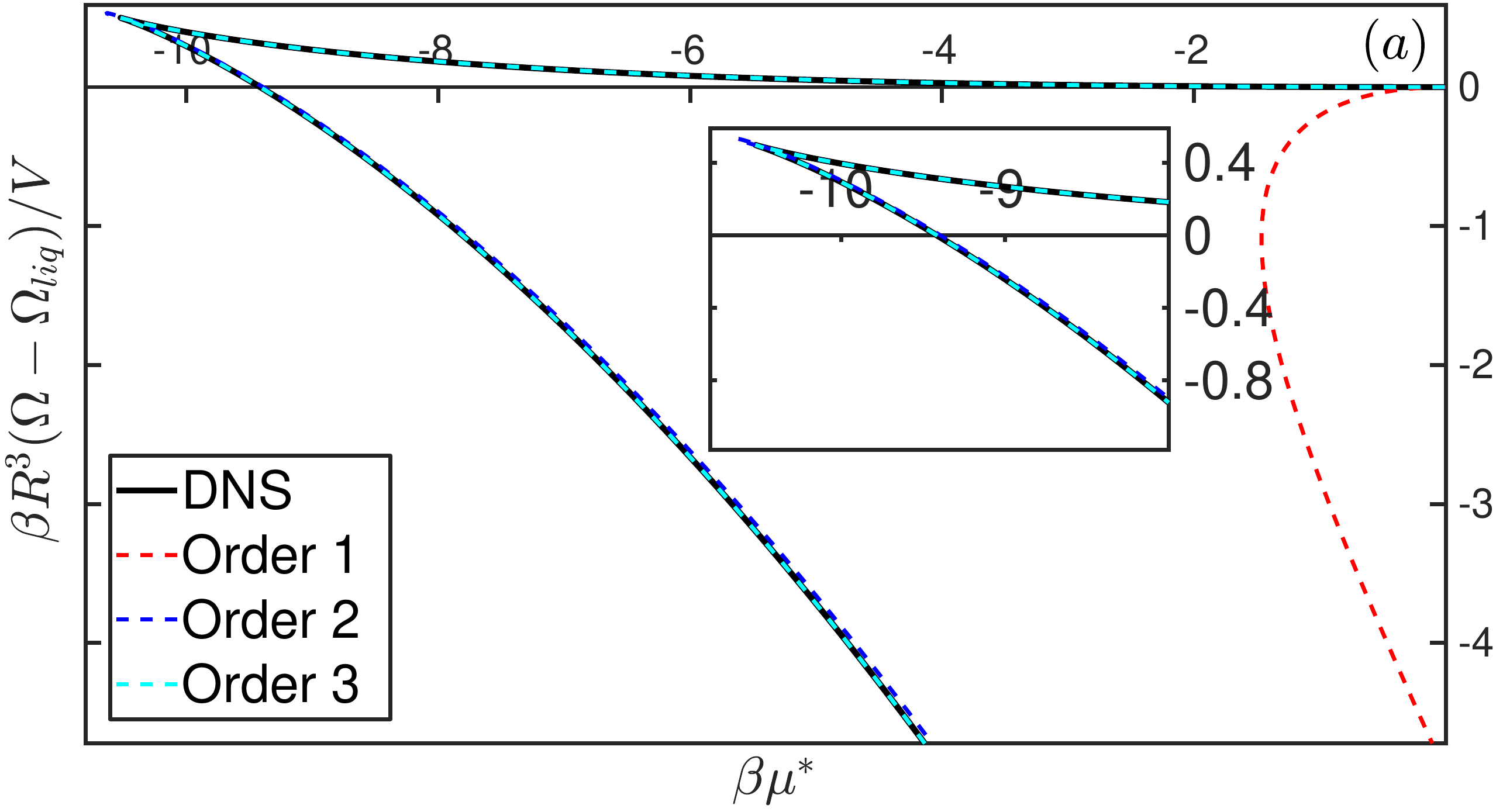}
\hspace{0.5cm}
\includegraphics[width=0.95\columnwidth]{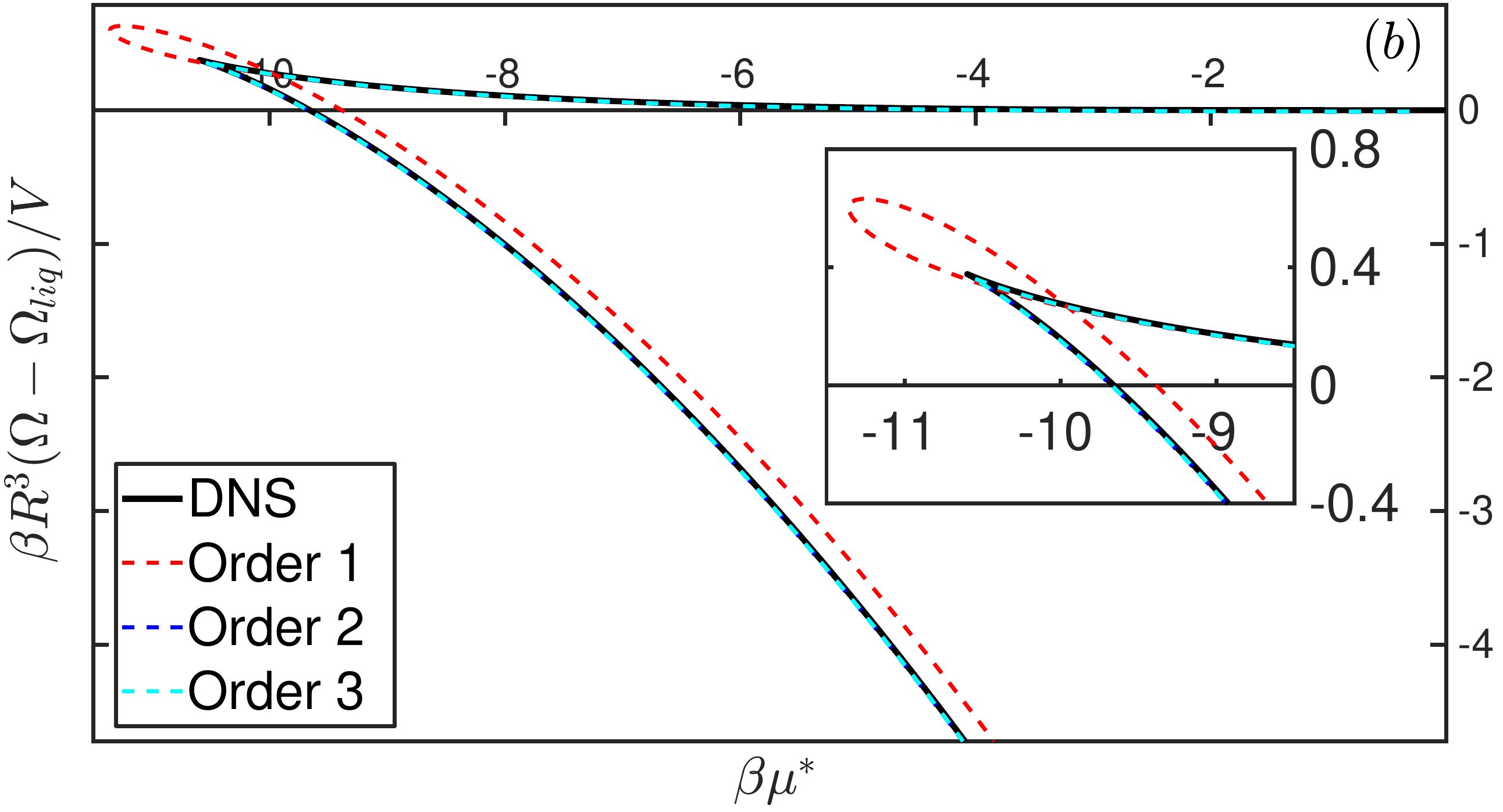}}
\caption{Specific grand potential $\beta R^3(\Omega-\Omega_{liq})/V$ (where $V$ 
is the volume) for $\beta\epsilon=1$ as a
function of the chemical potential~$\beta\mu^*$ for ($a$)~FCC crystals and
($b$)~BCC crystals in the GEM-4 model, at order~1, 2 and~3 in
SNLT, compared with fully resolved direct numerical solutions (DNS) of~(\ref{eq:eulerlagrangefinal}). {\color{black} All our results for crystal properties from order~3 SNLT are indistinguishable from the DNS.}}
 \label{fig:1}
 \end{figure*}
 
Here, our strongly nonlinear theory (SNLT) is a severe (but controlled) truncation
of~(\ref{eq:exp_sum_of_fourier}), along with the requirement that Fourier modes whose indices are
permutations of each other have equal amplitude. We refer to the level of truncation as the `order' of the~\hbox{SNLT}. In the supplementary material~\cite{suppmat} we give a detailed
exposition of SNLT and \textsc{Matlab} code applying it to the FCC crystal.
 

To illustrate the advantage of SNLT, 
we consider two different model systems
in~3D: (i) the generalized exponential
model with exponent 4 (GEM-4)~\cite{Mladek2006PRL, pini2015unconstrained}, which enables to compare our SNLT results with those of Ref.~\cite{pini2015unconstrained}, where an unconstrained minimization of $\Omega$ was performed, and (ii) a modified Barkan--Engel--Lifshitz (BEL)~\cite{barkan2014controlled} model designed to promote the formation of IQCs. For both we use the random phase approximation for the excess free energy
 \begin{equation}
 \cFex[\rho] = \frac{1}{2}\int\int 
                          \rho(\br) v(|\br-\br'|) \rho(\br')
                          \rdbr' \rdbr\,,
 \label{eq:fexRPA}
 \end{equation}                        
{\color{black}which} is accurate for soft-core systems \cite{likos2001effective}.
For particles with a hard-core, one should instead use an alternative approximation for $\cFex[\rho]$, e.g.\ one based on the highly accurate fundamental measure functionals for hard-spheres \cite{Evans92, hansen, roth2010fundamental}{\color{black}; see also \cite{denton1997thermodynamically_a, denton1997thermodynamically_b, denton1998stability, roth2000solid} for hard-core systems that form QCs.}
Taking~\eqref{eq:eulerlagrangephi} with~\eqref{eq:fexRPA}
yields                     
 \begin{equation}
 \phi(\br) + 
 \rho_0\beta \!\! \int \! v(|\br-\br'|) e^{\phi(\br')} \rdbr'  
 - \beta\mu^* = 0,
 \label{eq:eulerlagrangefinal}
 \end{equation}
recalling that $\exp(\phi)$ is proportional to the density.

The GEM-4 is a simple model of dendrimers in
solution, treating the effective interactions between the centers of mass via the pair potential $v(r)=\epsilon\exp(-r^4/R^4)$, with $\epsilon$ denoting the strength of the
interaction and $R$~its range. Figure~\ref{fig:1} shows the GEM-4 grand
potential minus that of the uniform liquid state per unit volume~($V$), $\beta R^3(\Omega-\Omega_{liq})/V$,
versus the chemical potential
$\beta\mu^*$ for successive orders of SNLT calculations for FCC and BCC crystals,
compared with full numerical solutions of Eq.~(\ref{eq:eulerlagrangefinal}) (an unconstrained minimization, using the approach described in~\cite{archer2019deriving}). We see that the order~1 SNLT approximation (red dashes) fails to
describe the crystal accurately, {\color{black} especially for the FCC,} but the order~2 and~3 SNLT perform
significantly better, to the extent that order~3 calculations (cyan dashes) overlap with the full numerical solutions (black solid line). Using this accurate order~3 SNLT, for $\beta\epsilon=1$ we find that the uniform liquid state transitions to a BCC phase at $\beta\mu^*=-9.67$, which itself then transitions to a FCC phase at $\beta\mu^*=-5.06$. The corresponding coexisting densities at the liquid--BCC transition are, $R^3{\bar\rho}_{\text{liq}}=5.55$ and $R^3{\bar\rho}_{\text{BCC}}=6.10$, while for the BCC--FCC transition we have, $R^3{\bar\rho}_{\text{BCC}}=7.65$, $R^3{\bar\rho}_{\text{FCC}}=7.70$. These SNLT values agree well with results from Pini et al.~\cite{pini2015unconstrained} and can easily be rescaled to obtain corresponding values at other temperatures \cite{archer2019deriving}.  Other periodic structures, such as lamellar, columnar hexagons and simple cubic crystals, are never global minima of the grand potential.
 
\begin{figure}[t]
\centering{
\includegraphics[width=0.85\columnwidth]{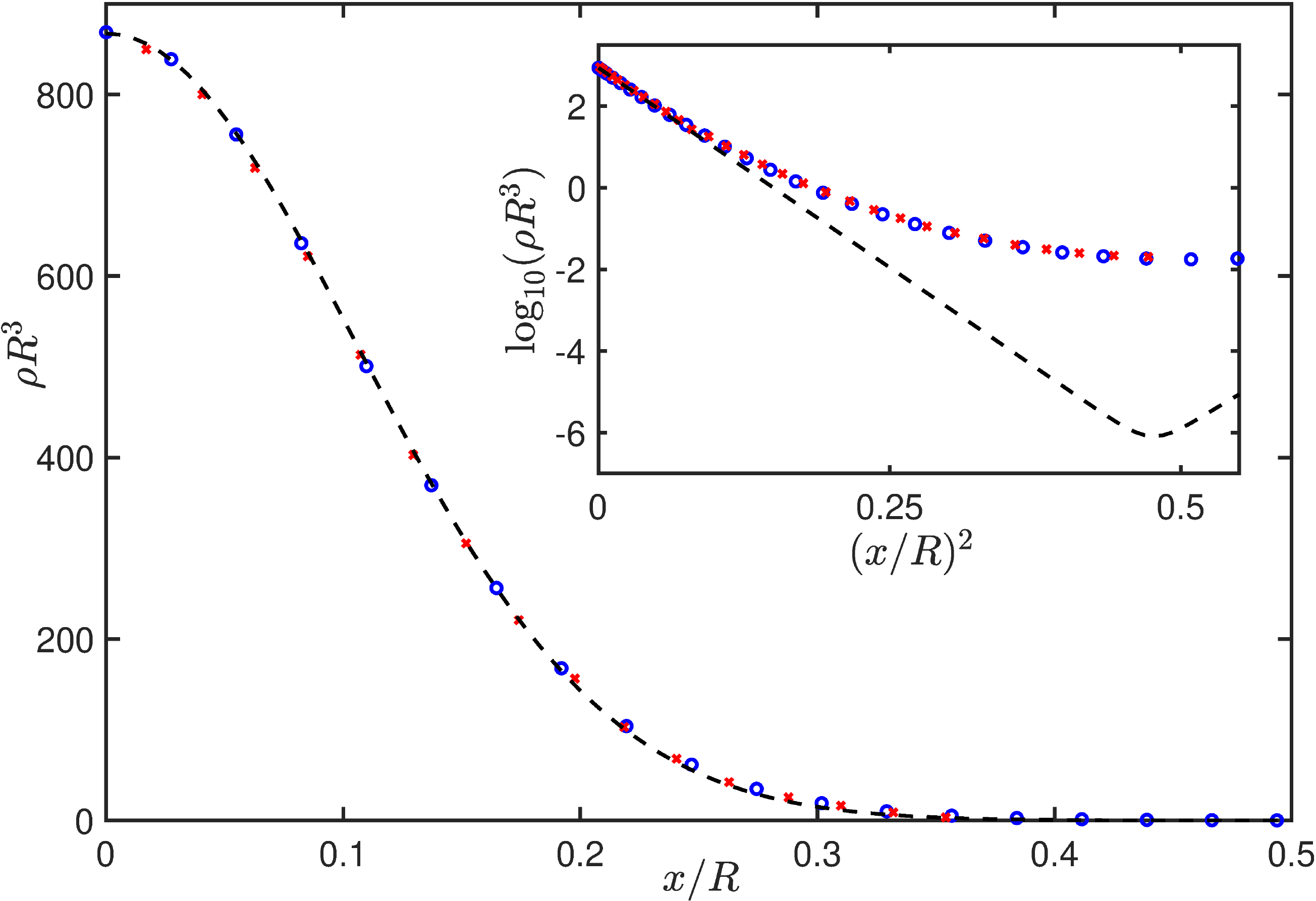}
}
\caption{\textcolor{black}{The density profile $\rho$ in a FCC crystal as a function of $x/R$, where $x$ is the distance along a path joining two nearest neighboring density peaks and $R$ is the range parameter in the pair potential, for $\beta\epsilon=1$ and $R^3\bar{\rho}=8.3$}. The red crosses are the unconstrained
minimization results from Figs.~2 and~3 in~\cite{pini2015unconstrained},
the blue circles are the order~3 SNLT results and the black dashed line is
the Gaussian form~\eqref{eq:sum_of_gaussians}. The inset shows
$\log_{10}(\rho R^3)$ as a function of~$(x/R)^2$, where~\eqref{eq:sum_of_gaussians} is a straight line.}
 \label{fig:2}
 \end{figure}


Figure~\ref{fig:2} shows the density distribution~$\rho$ as a function
of the interpeak distance $x$  
in the FCC crystal, calculated
from~\hbox{SNLT} (blue circles), from the unconstrained minimisation in
Figs.~2 and~3 of Ref.~\cite{pini2015unconstrained} (red crosses), and from
assuming the Gaussian form \eqref{eq:sum_of_gaussians} (dashed black line). Both SNLT and the Gaussian form~\eqref{eq:sum_of_gaussians}
match {\color{black} \cite{pini2015unconstrained}} well on the scale of the main plot. 
However, in the inset we plot $\log_{10}\rho$ as a function of~$x^2$, which
highlights the density between peaks, where we observe
that the results of~\cite{pini2015unconstrained} and SNLT {\color{black} both} deviate significantly from the Gaussian form.
This highlights an important weakness of
representation~(\ref{eq:sum_of_gaussians}): it underestimates the density
between peaks by several orders of magnitude. The density between
peaks gives the 
particle hopping rate between peaks, thus errors in calculating this leads to
errors in the diffusion coefficient and related transport properties.

\begin{figure}[]
\centering{
\includegraphics[width=0.85\columnwidth]{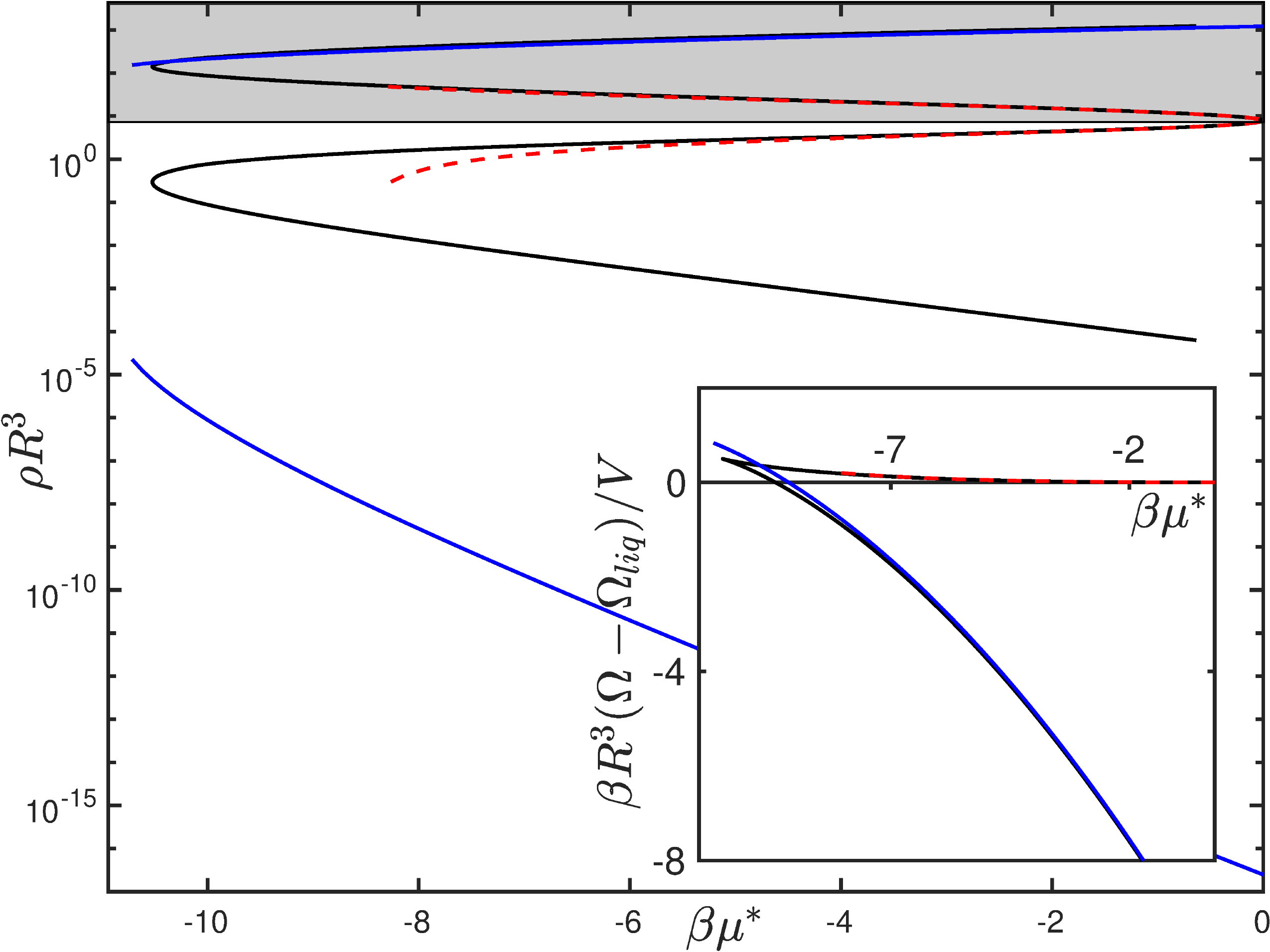}
}
\caption{Maximum (in region with grey background) and minimum (white background)
of $\rho$ as a function of~$\beta\mu^*$ for the
GEM-4 model in 3D, for FCC crystals. The inset shows the
grand potential of the crystal minus that of the liquid per unit volume. The
solid black lines are results from order~3 SNLT (including both the stable and unstable branches of solutions), red dashed lines
resulting from a crystal approximation method~\cite{jiang2020stability} that
uses Eq.~(\ref{eq:sum_of_fourier}), and the solid blue lines are obtained by using
Eq.~(\ref{eq:sum_of_gaussians}) to approximate the density distribution.}
 \label{fig:5}
\end{figure}

Figure~\ref{fig:5} shows the maximum and minimum of $\rho$ as a function
of~$\beta\mu^*$ obtained by three different methods, to compare the 
regimes under which the 
different representations of $\rho$ are valid.
The inset compares their grand potentials. The Gaussian representation~(\ref{eq:sum_of_gaussians})
(blue solid lines) recovers the maximum of the 
density profile correctly,
but underestimates the minimum significantly, in line with
Fig.~\ref{fig:2}. This form also leads to {\color{black} an overestimate in
the value of the grand potential, particularly near to melting}. 
The red dashed lines are results from the
crystal approximation method of Ref.~\cite{jiang2020stability} employing the
representation~(\ref{eq:sum_of_fourier}), also truncated at order~3. This gives the unstable lower solution branch well, but not the upper solution branch (going to much higher order is required to obtain the stable upper branch accurately \cite{archer2019deriving}). In contrast, the black solid line order~3 SNLT accurately captures the form of the density distribution for both branches, near and far away from melting.


Even though the density $\rho$ varies over many orders of
magnitude, fewer than a dozen independent Fourier amplitudes
in~\eqref{eq:exp_sum_of_fourier} are needed to represent it, while a
full Fourier representation~\eqref{eq:sum_of_fourier} requires
$\mathcal{O}(48^3)$ modes to resolve the peaks accurately. On the other hand, 
using sums of Gaussians~\eqref{eq:sum_of_gaussians} requires even fewer degrees
of freedom (only~$\alpha$, $n$ and $|\bR_l|$), but as Fig.~\ref{fig:5} shows, this
representation is less accurate close to melting, particularly
in determining the minimum of~$\rho$ and the grand potential.

The reason for such {\color{black}remarkable} efficiency of~\hbox{SNLT} is that the convolution in~\eqref{eq:eulerlagrangefinal} strongly damps modes with wavenumbers greater
than some cut-off value (which depends on the particular system), as pointed out in~\cite{archer2019deriving}. The density $\rho=\rho_0\exp(\phi)$ 
is sharply peaked and so has large amplitudes over
a wide range of Fourier modes, but when multiplied by~$v(r)$ and averaged in the convolution, high wavenumber modes are damped. Of the three terms
in~(\ref{eq:eulerlagrangefinal}), the last ($\beta\mu^*$) has only wavenumber zero, the second (convolution) term has only wavenumbers up to a cut-off, and so the first
term $\phi(\br)$ can also only contain wavenumbers up to the same cut-off, and so can be represented accurately with relatively few Fourier modes. {\color{black} Thus, the logarithm of the sharply peaked density is a smooth function.}

For the GEM-4 case, modes
with wavenumbers~$\gtrsim2/R$ are strongly damped~\cite{archer2019deriving}, and (for crystals) SNLT of order~4 or higher includes only modes with wavenumbers above this
cut-off (see the supplementary material \cite{suppmat}), so order~3 SNLT is sufficient.
The limited number of unknowns needed in SNLT makes it possible to determine crystal structures and compute phase diagrams using simple root finding packages (such as
\texttt{fsolve}) or minimization packages in \textsc{Matlab}. Since the exact Eq.~\eqref{eq:eulerlagrangephi} has a similar structure to the approximate
Eq.~\eqref{eq:eulerlagrangefinal} -- recall that all accurate DFTs are constructed from convolutions of the density with bounded functions (so-called weight functions)
\cite{Evans92, hansen, roth2010fundamental} -- therefore the above argument that SNLT is accurate for periodic crystals because Fourier modes in $\ln\rho$ above a certain
cut-off are strongly damped, applies in general, as long as the Fourier transform of {\color{black}the weight functions are short ranged. This is equivalent to the condition that the Fourier transform of $\int_0^\lambda \textrm{d}\lambda c^{(2)}(\br,\br';\rho_\lambda)$ [see \eqref{eq:exact}] becomes small beyond some cut-off. Thus, we expect SNLT to be widely applicable, not just to soft-core particles, although other systems may have the cut-off at larger $k$ than for GEM-4 model, requiring one to go a few orders higher for the SNLT to converge.}


The efficiency of the truncated SNLT for crystals relies on the fact that there
are a limited number of RLVs within the cut-off wavenumber.  In contrast, the
Fourier spectrum for QCs is dense~\cite{levine1984quasicrystals}, and there {\color{black} is}
an infinite number of Fourier modes within any cut-off sphere. Including more
modes in SNLT and/or using six-dimensional projection
methods~\cite{jiang2020stability,jiang2017stability,jiang2018icm} turns out
to be unsatisfactory because we get solutions only to a few
digits of accuracy. Nonetheless, these provide good approximate
initial conditions for other methods (such as Picard iteration used here), so we
still advocate using the representation~\eqref{eq:exp_sum_of_fourier}
and SNLT, combined with these other methods, for~QCs.

\begin{figure}[t]
\includegraphics[width=0.85\columnwidth]{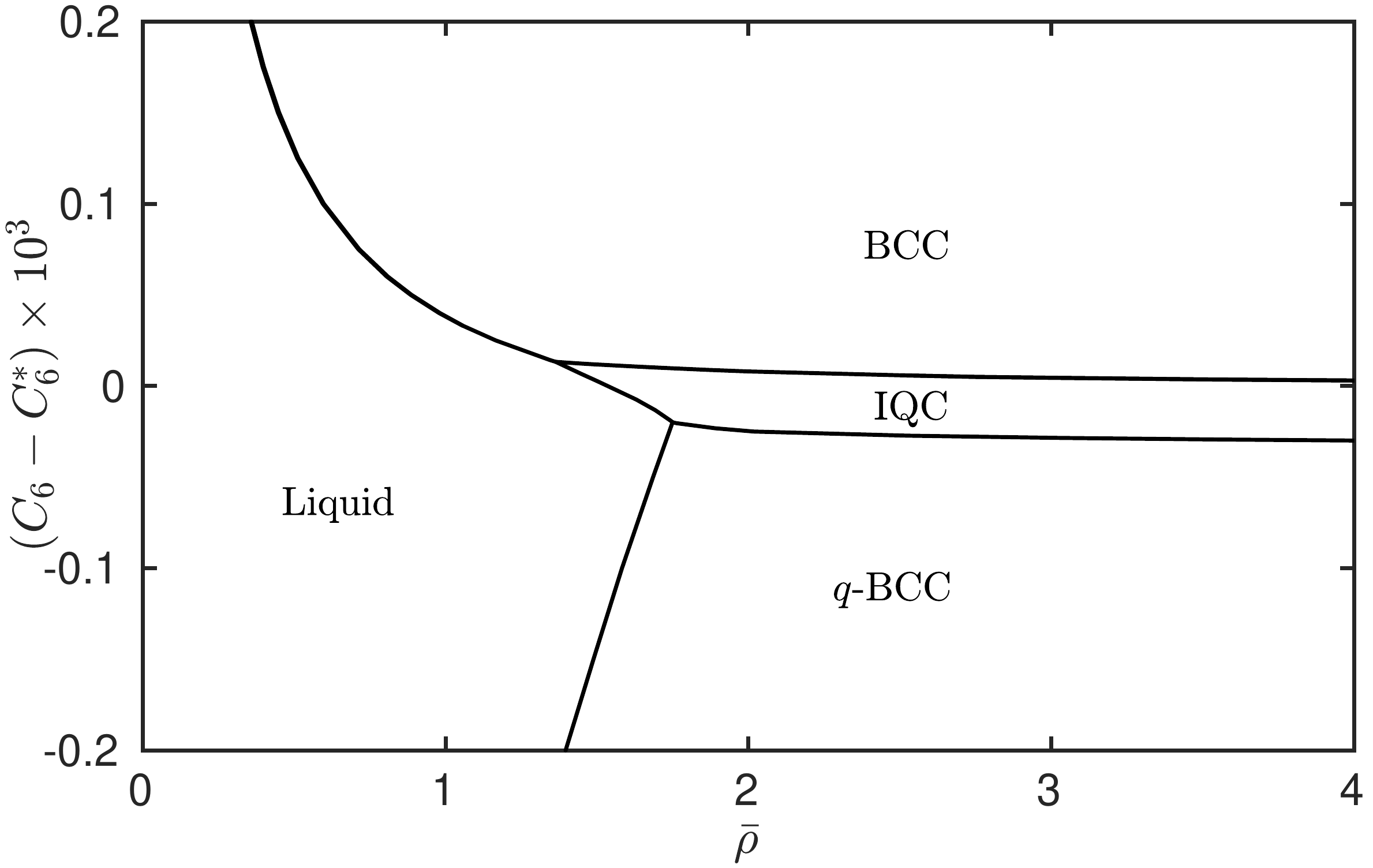}
\caption{Phase diagram for the BEL system (\ref{eq:BEL}) in the average density $\bar{\rho}$ versus $C_6$ pair potential parameter plane. The IQC arises
between the large lattice spacing BCC and small lattice spacing $q$-BCC. A liquid phase is also observed. The coexistence region between phases is of order the width of the lines.}
 \label{fig:4}
\end{figure}

We demonstrate this in a QC-forming system of soft particles interacting via the BEL 
pair potential~\cite{barkan2014controlled,ratliff2019wave}
 \begin{equation}\label{eq:BEL}
 v(r) = e^{-\frac{1}{2}\sigma^2 r^2}\sum_{n=0}^4 C_{2n}r^{2n}\,,
 \end{equation}
 which was previously shown to form QCs in 2D~\cite{barkan2014controlled,ratliff2019wave}. Here, we show that when the parameters $\{C_{2n},\sigma\}$ which control the form and range of $v(r)$ are chosen correctly, then this system also forms QCs in 3D.
 The values of $\{C_{2n},\sigma\}$ determine two characteristic lengthscales in the particle interactions, which we choose to be in the golden ratio $2\cos(\frac{\pi}{5})\approx1.618$, in order to encourage IQCs~\cite{subramanian2016three}. We choose $\sigma$ to promote IQC stability whilst keeping $v(r)\geq0$ for all $r$~\cite{ratliff2019wave}. Further details appear in the supplementary material \cite{suppmat}. To compute the phase diagram, we vary the coefficient~$C_6$ in~(\ref{eq:BEL}) and perform order~3 SNLT calculations for varying
$\beta\mu^*$ ($C_6^*$ denotes the value at which the system is exactly marginally unstable at the two lengthscales). This is sufficient to accurately determine the periodic crystalline phases. However, for the IQC phase, we use the order~3 SNLT result as an initial condition for a Picard iteration solver \cite{roth2010fundamental, hughes2014introduction}. 
Figure~\ref{fig:4} displays the resulting phase diagram, which exhibits the liquid and two BCC crystals. The $q$-prefix denotes the crystal with lattice spacing determined by the smaller characteristic lengthscale (larger wavenumber). In between these two, the IQC emerges as the minimum of the grand potential $\Omega$. In 
parts of the region considered, the FCC structure is a 
local minimizer, but is never the global minimum. {\color{black} We have not calculated the free energy for all possible structures, but of the likely candidates, the IQC is the global minimum in a portion of the phase diagram.}

Favourable contributions to $\Omega$ come from
triangles and pentagons (combinations of three or five wavevectors that add up to zero) in the spectrum of~$\rho$. Their abundance has been invoked to explain BCC~\citep{AlexanderMcTague} and QC stability~{\color{black} \citep{Bak1985, mermin1985mean, Lifshitz2007, roan1998stability, subramanian2016three}}. However, the sharp peaks in~$\rho$ and the consequential flatness of its spectrum obscures this argument. Our observation of strong damping at large $k$ in the spectrum of~$\ln\rho$ suggests that the triangle argument should be reframed in terms of this field. 

In summary, we have demonstrated that SNLT, representing $\ln\rho$
as a truncated Fourier sum~(\ref{eq:exp_sum_of_fourier}), is accurate
at all state points, both near and far from melting.
It is more efficient than representing $\rho$ as a
Fourier sum~(\ref{eq:sum_of_fourier}), and it has a wider range of validity
than representing it as a sum of Gaussians~(\ref{eq:sum_of_gaussians}),
which fails near melting {\color{black} and always predicts} the density to be too low
between the peaks. We expect SNLT to also be accurate for
bicontinous and similar phases exhibited by e.g.\ the
binary mixture considered in \cite{pini2015unconstrained}. For QCs, we
advocate SNLT as a method of generating good starting profiles for other
(iterative) methods. Even without the SNLT severe truncation,
in all cases we expect
representation~(\ref{eq:exp_sum_of_fourier}) to be superior
to~(\ref{eq:sum_of_fourier}).

\begin{acknowledgments}
This work was supported by a Hooke Research Fellowship (PS), 
the EPSRC under grants EP/P015689/1 (AJA, DJR) and
\hbox{EP/P015611/1 (AMR)}, and the Leverhulme Trust (RF-2018-449/9, AMR).
This work was undertaken on ARC4, part of the High Performance Computing facilities at the University of Leeds, UK.
We acknowledge Ken Elder and Joe Firth for valuable discussions.
\end{acknowledgments}



%


\end{document}